\documentclass[conference]{IEEEtran}
\IEEEoverridecommandlockouts
\usepackage{cite}
\usepackage{algorithmic}
\usepackage{graphicx}
\usepackage{textcomp}
\usepackage{xcolor}
\usepackage{xspace}
\usepackage{url}
\usepackage{amsmath}
\usepackage{multicol}
\usepackage[utf8]{inputenc}
\usepackage{amsmath}
\usepackage{amsfonts}
\usepackage{amssymb}
\usepackage{multirow}
\usepackage{tabularx}
\usepackage{bm}

\def\BibTeX{{\rm B\kern-.05em{\sc i\kern-.025em b}\kern-.08em
    T\kern-.1667em\lower.7ex\hbox{E}\kern-.125emX}}

\usepackage{pythonhighlight}

\newcommand{\vect}[1]{{#1}}
\newcommand{\toolname}{\texttt{PortFawn} \xspace}
\newcommand{\toolnamens}{\texttt{PortFawn}}

\newcommand{\DateStart}{January 1, 2015 \xspace}
\newcommand{\DateEnd}{December 31, 2020 \xspace}
\newcommand{\RiskFreeRate}{0.0\% \xspace}
\newcommand{\ParamR}{20.0\% \xspace}
\newcommand{\ParamV}{5.0\% \xspace}
\newcommand{\TrainDaysNum}{40 \xspace}
\newcommand{\TestDaysNum}{5 \xspace}
\newcommand{\AssetNum}{nine \xspace}
\newcommand{\AssetMinW}{2.0\% \xspace}
\newcommand{\AssetMaxW}{98.0\% \xspace}
\begin{document}

\title{Portfolio Optimization \\ on Classical and Quantum Computers Using \toolname
\thanks{\toolname is an open-source Python package under an MIT license to provide easy-to-use and well-tested tools to researchers. \toolname should solely be used for educational and research purposes and not as an investment analysis tool. Individuals who seek financial and investment data, information, visualizations, and advice are highly encouraged to seek personalized advice from a certified financial advisor. 
The authors do not accept any responsibility and/or liability for the consequences or any damages resulting in using this tool for any purposes other than research and education.
We use the D-Wave quantum computer (\textit{Developer} account) and Ocean software infrastructure as part of our evaluations.}
}

\author{
\IEEEauthorblockN{Moein Owhadi-Kareshk}
\IEEEauthorblockA{\textit{Department of Computing Science} \\
\textit{University of Alberta}\\
Edmonton, Canada \\
owhadika@ualberta.ca}
\and
\IEEEauthorblockN{Pierre Boulanger}
\IEEEauthorblockA{\textit{Department of Computing Science} \\
\textit{University of Alberta}\\
Edmonton, Canada \\
pierreb@ualberta.ca}
}

\maketitle

\begin{abstract}
Portfolio diversification is one of the most effective ways to minimize investment risk. Individuals and fund managers aim to create a portfolio of assets that not only have high returns but are also uncorrelated. This goal can be achieved by comparing the historical performance, fundamentals, predictions, news sentiment, and many other parameters that can affect the portfolio's value.
One of the most well-known approaches to manage/optimize portfolios is the well-known mean-variance (Markowitz) portfolio. The algorithm's inputs are the expected returns and risks (volatility), and its output is the optimized weights for each asset in the target portfolio. Simplified unrealistic assumptions and constraints were used in its original version preventing its use in practical cases. One solution to improve its usability is by altering the parameters and constraints to match investment goals and requirements. This paper introduces \toolname, an open-source Python library to create and backtest mean-variance portfolios. \toolname provides simple-to-use APIs to create and evaluate mean-variance optimization algorithms using classical computing (real-valued asset weights) as well as quantum annealing computing (binary asset weights). This tool has many parameters to customize the target portfolios according to the investment goals. The paper introduces the background and limitations of the mean-variance portfolio optimization algorithm, its architecture, and a description of the functionalities of \toolname. We also show how one can use this tool in practice using a simple investment scenario.
\end{abstract}


\section{Introduction} \label{sec:introduction}

There is currently a vast range of tools for investors to help invest in equities, fixed-income, real estates, commodities, futures, etc. Each of these investment tools has its characteristics, such as risks and returns. 
Investors usually choose one or multiple of them to make an investment based on their financial needs. 
It is shown that investing in assets with uncorrelated returns might help at decreasing risks while only slightly sacrificing on returns~\cite{markowitz1959portfolio}. Investors, thus, create a portfolio that is a basket of different assets to achieve their financial goals while eliminating unnecessary risks.

There are two main types of risks for each asset, systematic and non-systematic risks~\cite{dowen1988beta, cheng1997switching, sofrankova2017actual, todorov2010jumps}. 
The systematic risks are common to all or most assets, while the non-systematic risk only corresponds to a specific asset. As an example of systematic risk, if the Gross Domestic Product (GDP) of a country dramatically decreases due to poor monetary policies, all of the assets listed in the markets will be impacted. This is systematic risk, and investors who invest in such a country cannot decrease the risk. 
However, a decrease in the return of just one company in the market due to bad business decisions is an example of non-systematic risk, and investors can decrease its impact on their portfolio by diversification. 
A good portfolio aims at holding assets with high returns that are diverse and uncorrelated.

In this paper, we focus on the mean-variance portfolio optimization approach, which is based on the Markowitz theory~\cite{markowitz1959portfolio}. 
This approach is based on a sound theoretical background that uses the expected (mean) returns and risks (the covariance matrix of asset returns) to minimize the risk of a portfolio for a given target return, or equivalently, maximize the portfolio return for a target risk level. 
This approach defines portfolio optimization as a quadratic (or linear) programming problem with a set of linear (or quadratic) constraints. Summarizing the returns data to the expected returns and the covariance matrix, and eliminating other available information about assets can be seen as an oversimplification. 
However, this theory is effective enough in some scenarios and can be used as a complementary approach to other investment approaches. 

The mean-variance portfolio is vastly studied in the literature, including studies on employing different optimization methods such as linear models~\cite{mansini2014twenty} and evolutionary optimization methods~\cite{metaxiotis2012multiobjective, ponsich2012survey}.
The mean-variance portfolio optimization can be expressed as either single~\cite{aranha2009memetic, corazza2013particle, strumberger2016constrained} or multi-objective optimization problem~\cite{mishra2002fast, armananzas2005multiobjective, baykasouglu2015grasp, anagnostopoulos2011mean, fliege2014robust, lwin2014learning, shaverdi2020multi, garcia2020multiobjective}. 
Adding additional constraints to the original mean-variance portfolio optimization algorithm is also investigated in the literature to make the assumptions and constraints more realistic~\cite{krokhmal2002portfolio, crama2003simulated, peng2011new, soleimani2009markowitz, suganya2009constrained, meghwani2017multi, babazadeh2019novel}.

This paper introduces \toolname, an open-source Python library that we have developed to implement and backtest the mean-variance portfolio optimization approach with options to change the optimization goal, constraints, and other portfolio parameters. 
This tool (1) collects the asset prices for a given list of assets, (2) calculates the expected returns and risks, (3) builds a portfolio, and (4) performs backtesting to evaluate the optimized portfolio. 
\toolname supports multiple cost functions and constraints where each can be used under certain assumptions. 
Besides the ability to run the optimization on classical computers, \toolname also solves the portfolio optimization problem using quantum annealing computers. 

This is a meta-heuristic optimization approach that leverages the power of quantum computers to optimize a given objective function with binary variables~\cite{kadowaki1998quantum, santoro2002theory, boixo2014evidence}. 
This approach has a close relation to its classical version, simulated annealing~\cite{kirkpatrick1983optimization, van1987simulated, finnila1994quantum}. 
Quantum annealing is based on adiabatic quantum computing capable of solving Quadratic Unconstrained Binary Optimization (QUBO). The solver aims to minimize the energy Hamiltonian (function) of the optimization problem by slowly applying the problem Hamiltonian to an initial quantum state~\cite{childs2001robustness, mcgeoch2014adiabatic, albash2018adiabatic}. 
Optimization by quantum annealing has been successfully applied to a vast range of problems in the literature such as maximum-cut~\cite{guerreschi2019qaoa, gharibian2019almost}, graph coloring~\cite{titiloye2011quantum, kudo2018constrained, silva2020mapping}, graph partitioning~\cite{ushijima2017graph, negre2020detecting}, independent set~\cite{yarkoni2018first}, knapsack~\cite{haddar2016hybrid}, satisfactory~\cite{battaglia2005optimization, hsu2018quantum, kruger2020quantum}, and machine learning~\cite{adachi2015application, crawford2016reinforcement, amin2018quantum} problems. 

Quantum annealing is also used to solve the portfolio optimization problem~\cite{venturelli2019reverse, cohen2020portfolio, grant2021benchmarking} before. Still, most of the literature only focuses on quantum computers' internal abstract performance and does not study the implication of using it in portfolio optimization. 
Moreover, developing a new problem formulation method needs individual efforts by researchers to develop the code for optimizing portfolios and backtesting them. \toolname provides a set of unified APIs to implement and evaluate any novel idea in this domain with minimum effort.

\toolname logs every step in building and backtesting portfolios to make the result validation and verification easier. It also contains unit and integration tests to ensure that the tool does not break after a change. 
We release the \toolname code with its documentation as an open-source tool under the MIT License\footnote{\url{https://github.com/mkareshk/PortFawn}}. We encourage the reader to read the online documentation for any recent changes and use this paper as a first step of understanding the background of the mean-variance portfolio optimization and general information about \toolnamens. We also invite the reader to contribute to the code and the documentation of this work.

We designed a portfolio optimization and backtesting scenario and show the sample results to demonstrate our tool's functionality. 
We use the historical data of \AssetNum Exchange-Traded Funds (ETFs) between \DateStart and \DateEnd for this backtesting. 
We use the data of \TrainDaysNum days for creating portfolios and \TestDaysNum for out-of-sample evaluation using overlapping time windows. The results show that each of the mean-variance portfolio optimizations works according to what we expected.
For example, we observe very low volatility in the minimum volatility portfolio and a high return in the maximum return portfolio. 

In Section~\ref{sec:portfolio} we explain how the mean-variance portfolio optimization works. We then introduce our \toolnamens, its architecture, and characteristics in Section~\ref{sec:architecture}. 
In Section~\ref{sec:backtesting} we perform backtesting using \toolname to show the functionality of our tool in practice.
We finally discuss the limitations of mean-variance portfolio optimization in Section~\ref{sec:limitatons}.

\section{Portfolio Optimization} \label{sec:portfolio}

It is important to understand how the mean-variance portfolio optimization works before explaining the main features of \toolnamens. 
In this section, we first review the calculations of returns and risks for assets and portfolios in Section~\ref{subsec:ReturnRiskAssets} and \ref{subsec:ReturnRiskPortfolios}, respectively. 
We then review the essential definitions of portfolio optimization in Section~\ref{subsec:PortfolioOptimization}. 
In the end, we explain how to use quantum annealing for portfolio optimization in Section~\ref{subsec:QuantumAnnealing}

\subsection{Returns and Risks of Assets} \label{subsec:ReturnRiskAssets}
The mean-variance portfolio optimization assumes that asset returns are i.i.d random variables derived from a normal distribution and can uniquely be parametrized by their means and variances. 
If the price of asset $i$ changes from $p_1$ to $p_2$ during an specific period of time, the asset return $r_i$ is calculated as

\begin{equation}
r_i = \frac{p_2-p_1}{p_1}.\label{eq:AssetReturn}
\end{equation}

In reality, one can use \eqref{eq:AssetReturn} to calculate the returns only at the end of the investment period when both $p_1$ and $p_2$ are known. 
We, therefore, use the \textit{expected returns} in calculations to be able to create a portfolio for the future and analyze its performance. 
The expected return of an asset is a prediction (an educated guess) about the asset's performance in the future using historical data on price changes and/or fundamental asset information. 
However, assuming a normal distribution for the returns, using the average of the returns is the best guess and a natural choice as the expected return. 
If $r_i^1, r_i^2, ..., r_i^k$ are the returns of asset $i$ in $k$ subsequent historical observations, e.g. daily prices, the expected return is calculated by

\begin{equation}
\bar{r}_i=\mu_{r_i}=E[r_i]=\frac{1}{k}\sum_{j=1}^{k}r_i^j.\label{eq:AssetExpectedReturn}
\end{equation}

By considering volatility as a risk indicator, the variance is used to characterize the risk of asset $i$.

\begin{equation}
\sigma_i^2=E[r_i-\bar{r}_i]=\frac{1}{k-1}\sum_{j=1}^{k}(r_i^j-\bar{r_i})^2 \label{eq:AssetVariance}.
\end{equation}

However, the standard deviation is usually used in analysis since its unit is as same as the returns.

\subsection{Returns and Risks of a Portfolio} \label{subsec:ReturnRiskPortfolios}

Portfolio optimization aims to create an optimal basket of $N$ assets to reduce non-systematic (diversifiable) risks while gaining the highest possible return. 
Let $t_0$ to $t_1$ be the time frame to use its historical data to create a portfolio. 
We then hold the portfolio from time $t_1$ to $t_2$ and then evaluate its performance in this investment configuration. 
The mean and standard deviation of historical returns between $t_0$ and $t_1$ is the expected return and risk for $t_1$ to $t_2$. 
This is a very strong generalization that directly impacts portfolio performance.

We show the asset allocation weights by vector $\vect{w}=[w_1, w_2, ..., w_N]^T$ where $w_i \in \mathbb{R}$ is the weight of the $i^{th}$ asset in the portfolio such that $0 \leq w_i$ which means that we just allow long positions and avoid short-selling. 
Furthermore, we aim to invest all the available capital and thus, $\vect{w}^T\vect{1}=1$, i.e. $\sum_{i=1}^{N}w_i=1$. The vector representation of asset returns is $\vect{r}=[r_1, r_2, ..., r_N]^T$ and $\vect{\Sigma}$ is the covariance matrix of the asset returns.
The portfolio return and variance are calculated by 

\begin{equation}
r_p=\vect{w}^T\vect{r}=\sum_{i=1}^{N}w_ir_i \label{eq:PortfolioReturn},
\end{equation}

\begin{equation}
\sigma_p^2=\vect{w}^T\vect{\Sigma}\vect{w}=\sum_{i=1}^{N}\sum_{j=1}^{N} w_i\sigma_{ij}w_j \label{eq:PortfolioVariance}.
\end{equation}

\subsection{Mean-Variance Portfolio Optimization} \label{subsec:PortfolioOptimization} 

Now that we know how to calculate the risk and return of a portfolio, we can start finding the optimal asset allocation $\vect{w^*}$ for various objectives. 
In the rest of this subsection, we introduce different approaches to formulate the mean-variance portfolio optimization problem. 
Each of these approaches aims to find $\vect{w^*}$ that optimizes a specific objective concerning a set of constraints. 
Investment requirements and goals determine which approach to choose. 

\paragraph{Minimum Volatility Portfolio (MVP)}
The goal of this portfolio is to obtain the desired return $R$ while minimizing the volatility. 
It also ensures to invest the whole capital and do not perform any short-selling. 

\begin{subequations}
\begin{alignat}{2}
&\!\min_\vect{w}     &\qquad& \vect{w}^T\vect{\Sigma}\vect{w}                    \label{eq:OptimizationMRA} \\
&\text{subject to} &      &  \vect{w}^T{r}=R                                         \label{eq:OptimizationMRB}\\
&                  &      &  \vect{w}\vect{1}=1                                           \label{eq:OptimizationMRC}\\
&                  &      &   w_i \geq 0, i=1, ..., N  \label{eq:OptimizationMRD}
\end{alignat}
\end{subequations}

\paragraph{Maximum Return Portfolio (MRP)}
This approach aims to maximize the portfolio's returns for a given desired volatility $V$. 
Similar to MVP, this approach also ensures investing all available capital and does not perform short-selling. 

\begin{subequations}
\begin{alignat}{2}
&\!\max_\vect{w}         &\qquad& \vect{w}^T\vect{r} \label{eq:OptimizationMREA} \\
&\text{subject to} &      & \vect{w}^T\vect{\Sigma}\vect{w}=V  \label{eq:OptimizationMREB}\\
&                  &      & \vect{w}\vect{1}=1  \label{eq:OptimizationMREC}\\
&                  &      & w_i \geq 0, i=1, ..., N \label{eq:OptimizationMRED}
\end{alignat}
\end{subequations}

\paragraph{Maximum Sharpe Ratio Portfolio (MSRP)}
The Sharpe ratio is one of the risk-adjusted return indicators based on the risk-free asset return $r_{rf}$.

\begin{equation}
S_p = \frac{r_p - r_{rf}}{\sigma_p}. \label{eq:SharpeRatio}
\end{equation}

The risk-free asset is considered a safe investment with zero risks. 
While no asset can be regarded as entirely safe, a set of them is assumed to be risk-free in quantitative analysis. 
Treasury bills issued by the U.S. Department of the Treasury~\footnote{https://www.treasury.gov/} is the one that is widely used in the industry. 

This approach aims to maximize $S_p$, while ensuring investing all available capital and using long positions only: 

\begin{subequations}
\begin{alignat}{2}
&\!\max_\vect{w}         &\qquad& S_p \label{eq:OptimizationMSA} \\
&\text{subject to} &      & \vect{w}\vect{1}=1 \label{eq:OptimizationMSB}\\
&                  &      & w_i \geq 0, i=1, ..., N \label{eq:OptimizationMSC}
\end{alignat}
\end{subequations}

\paragraph{Multi-Objective Portfolio (MOP)}
This approach aims to optimize both portfolio return and volatility at the same time while ensuring investing all available capital with long positions only. The cost function of this portfolio is the linear combination of return and risk using the risk aversion parameter $\lambda$.

\begin{subequations}
\begin{alignat}{2}
&\!\min_\vect{w}         &\qquad& \vect{w}^T\vect{\Sigma}\vect{w} - \lambda \vect{w}^T\vect{r} \label{eq:OptimizationMVA} \\
&\text{subject to} &      & \vect{w}\vect{1}=1  \label{eq:OptimizationMVB}\\
&                  &      & w_i \geq 0, i=1, ..., N \label{eq:OptimizationMVC}
\end{alignat}
\end{subequations}

\paragraph{Market Capitalization Portfolio (MCP)}
Contrary to the previous portfolio optimization approaches, this approach does not need to solve an optimization problem. Instead, it assigns weights to each asset according to its market capitalization. Most passive index funds use this approach for asset management.

\paragraph{Equal-Weighted Portfolio (EWP)}
As the name suggests, this portfolio assigns equal weights to all assets regardless of any other factor. If the investment universe is diverse (as in our experiment in this paper), this approach can be pretty practical, bringing simplicity to investment.

\subsection{Portfolio Optimization Using Quantum Annealing} \label{subsec:QuantumAnnealing}
To use quantum annealing, the objective function of the portfolio optimization problem is represented as a QUBO model and the aim is to \textit{select} or \textit{not select} assets to create a long-only portfolio. Therefore, the asset weights are binary, and not real-valued here.

In a QUBO model with $N$ binary variables, each variable is represented by $x_i\in\{0, 1\}$, its linear bias is $Q_{ii}$ and its quadratic bias with an arbitrary variable $x_j$ is $Q_{ij}$. A QUBO model can be also represented as a graph where $E=\{x_1, x_2, ..., x_N\}$ is the set of nodes and $Q$ is the adjacency matrix. The QUBO model defines the optimization objective as an energy function.

\begin{equation}
E(x) = \sum_{i=1}^{N}Q_{ii}x_i + \sum_{\substack{i, j=1 \\ i<j}}^{N}Q_{ij}x_ix_j. \label{eq:QUBOModel}
\end{equation}

The scalar notation of MOP's objective \eqref{eq:OptimizationMVA} is 

\begin{equation}
\sum_{i=1}^{N}-\lambda r_iw_i + \sum_{i, j=1}^{N}\sigma_{ij}w_iw_j \label{eq:MOPSclaar},
\end{equation}

which is identical to the QUBO model \eqref{eq:QUBOModel} when $x_i=w_r$, $Q_{ii}=-\lambda r_i$, and $Q_{ij}=\sigma_{ij}$. We use QUBO to program a multi-objective portfolio optimization problem ($\lambda=1$ for simplicity) on a quantum computer and call it Binary Multi-Objective Portfolio (BMOP). We ensure the two constraints (long-only positions and investing all available capital) by normalizing the wights. 

Finding a local minimum (solution), instead of the global minima, is a common challenge in classical solvers. However, quantum solvers have a higher chance of finding the global minimum by utilizing the quantum phenomena called \textit{quantum tunnelling}.

\section{The Architecture of \toolname} \label{sec:architecture}
This section discusses the most important classes implemented in \toolname. 

\paragraph*{Historical Market Data} The \texttt{MarketData} class collects the historical prices for a given list of assets (tickers) and a time period using the \texttt{yfinance}\footnote{https://pypi.org/project/yfinance/} Python package and calculates basic statistics. This class caches the price data to make the external \texttt{yfinance} API calls efficient. 
Asset returns, cumulative returns, correlation and covariance matrices, and return distributions are calculated in this class. 
This class also has APIs to visualize the market data.

\paragraph*{Expected Returns and Risks} The \texttt{ExpectedStats} class calculates the expected returns and risks for a given \texttt{MarketData} instance. The mean-variance portfolio is susceptible to these two inputs, and therefore, it is important to make these robust as much as possible. Our tool supports three types of calculations for the expected returns and risks, as follows:

\begin{itemize}
    \item \textbf{full:} Simply calculates the mean and covariance of the returns using the whole data.
    \item \textbf{random:} Selects a subset of returns (a time window) randomly and calculates the mean and variance for only that subset, and repeats this process for a given number of times. The results are aggregated at the end using the median.
    \item \textbf{weighted:} Similar to the random option, but assign more weights to the most recent returns.  
\end{itemize}

\paragraph*{Portfolio} The \texttt{Portfolio} class is one of the essential classes in \toolname, which generates a portfolio for a given list of assets, period, and portfolio objective and constraints. 
This class also calculates and visualizes the performance metrics such as returns, risks, annualized returns, annualized risks, cumulative returns, and Sharpe ratio for the optimized portfolio. 
The portfolio optimization objectives currently supported by this class are equal-weighted, minimum risk, maximum return, and maximum Sharpe ratio. 
This class uses the \texttt{PortfolioOptimization} class to perform optimization on classical or quantum computers.

\paragraph*{Backtesting} Creating a portfolio for just a one-time window and testing it with a subsequent time window is a naive way of performance evaluation since the effectiveness of an investment strategy might change over time due to the changes in asset's fundamentals or macroeconomic factors.
For a meaningful evaluation, we need to create and test portfolios in multiple overlapping rolling windows to ensure the consistency of the results over time. 
The \texttt{BackTest} class performs backtesting in \toolname, and the user can choose the number of training and testing days, the period of backtesting, and all other portfolio parameters.

\section{Illustrative Example of Using \toolname in Practice} \label{sec:backtesting}
In this section, we demonstrate a show-case for using \toolname. We do not explicitly mention the code to generate figures since the \toolname API might change in the future. Still, the examples with the updated code are available on the documentation page.

\subsection{Backtesting Setup}
First, let's define the assumptions and parameters for our backtesting scenario that compares the available portfolio optimization approaches in \toolname. Table~\ref{tab:backtestingparams} introduces the parameters we choose for our experiment. We choose \AssetNum diverse iShares ETFs (listed in the U.S.) in Table~\ref{tab:assetlist} to be used in this experiment.

\begin{table}[!t]
\centering
\caption{Backtesting Parameters for Our Illustrative Example}
\label{tab:backtestingparams}
\begin{tabular}{|l|l|l|}
\hline
\textbf{\#} & \textbf{Parameter}        & \textbf{Value}    \\ \hline \hline
1           & Backtesting starting date    & \DateStart   \\ \hline
2           & Backtesting ending date      & \DateEnd \\ \hline
3           & Portfolio training period & \TrainDaysNum days          \\ \hline
4           & Portfolio testing period  & \TestDaysNum days           \\ \hline
5           & Risk-free rate            & \RiskFreeRate (for simplicity)            \\ \hline
6           & Returns to use            & Daily             \\ \hline
7           & Target return ($R$ in \eqref{eq:OptimizationMRB})           & \ParamR          \\ \hline
8           & Target volatility ($V$ in \eqref{eq:OptimizationMREB})            & \ParamV             \\ \hline
9           & Lower Bound Asset Weight  & \AssetMinW           \\ \hline
10           &Upper Bound Asset Weight  & \AssetMaxW           \\ \hline

\end{tabular}
\end{table}

\begin{table*}[!t]
\centering
\caption{List of Assets Used in Our Backtesting Scenario}
\label{tab:assetlist}
\begin{tabular}{|l|l|l|l|l|l|}
\hline
\textbf{\#}                      & \textbf{Ticker}           & \textbf{Asset Name}                                              & \textbf{Short Name}             & \textbf{Asset Class}              & \textbf{Market}               \\ \hline
\multicolumn{1}{|l|}{\textbf{1}} & \multicolumn{1}{l|}{IVV}  & \multicolumn{1}{l|}{iShares Core S\&P 500 ETF}                   & \multicolumn{1}{l|}{S\&P 500}   & \multicolumn{1}{l|}{Equity}       & \multicolumn{1}{l|}{U.S.}     \\ \hline
\multicolumn{1}{|l|}{\textbf{2}} & \multicolumn{1}{l|}{IJR}  & \multicolumn{1}{l|}{iShares Core S\&P Small-Cap ETF}             & \multicolumn{1}{l|}{Small-Cap}  & \multicolumn{1}{l|}{Equity}       & \multicolumn{1}{l|}{U.S.}     \\ \hline
\multicolumn{1}{|l|}{\textbf{3}} & \multicolumn{1}{l|}{ACWX} & \multicolumn{1}{l|}{iShares MSCI ACWI ex U.S. ETF}               & \multicolumn{1}{l|}{ACWI}       & \multicolumn{1}{l|}{Equity}       & \multicolumn{1}{l|}{Global}   \\ \hline
\multicolumn{1}{|l|}{\textbf{4}} & \multicolumn{1}{l|}{IEMG} & \multicolumn{1}{l|}{iShares Core MSCI Emerging Markets ETF}      & \multicolumn{1}{l|}{Emerging}   & \multicolumn{1}{l|}{Equity}       & \multicolumn{1}{l|}{Emerging} \\ \hline
\multicolumn{1}{|l|}{\textbf{5}} & \multicolumn{1}{l|}{REET} & \multicolumn{1}{l|}{iShares Global REIT ETF}                     & \multicolumn{1}{l|}{Global RS}  & \multicolumn{1}{l|}{REIT}         & \multicolumn{1}{l|}{Global}   \\ \hline
\multicolumn{1}{|l|}{\textbf{6}} & \multicolumn{1}{l|}{IYR}  & \multicolumn{1}{l|}{iShares U.S. Real Estate ETF}                & \multicolumn{1}{l|}{U.S. RS}    & \multicolumn{1}{l|}{REIT}         & \multicolumn{1}{l|}{U.S.}     \\ \hline
\multicolumn{1}{|l|}{\textbf{7}} & \multicolumn{1}{l|}{HYG}  & \multicolumn{1}{l|}{iShares iBoxx High Yield Corporate Bond ETF} & \multicolumn{1}{l|}{Corp. Bond} & \multicolumn{1}{l|}{Fixed-Income} & \multicolumn{1}{l|}{U.S.}     \\ \hline
\multicolumn{1}{|l|}{\textbf{8}} & \multicolumn{1}{l|}{AGG}  & \multicolumn{1}{l|}{iShares Core U.S. Aggregate Bond ETF}        & \multicolumn{1}{l|}{Agg. Bond}  & \multicolumn{1}{l|}{Fixed-Income} & \multicolumn{1}{l|}{U.S.}     \\ \hline
\textbf{9}                       & IAU                       & iShares Gold Trust                                               & Gold                            & Commodity                         & U.S.                          \\ \hline
\end{tabular}
\end{table*}

\subsection{Historical Market Data}

\begin{figure*}[!t]
    \centering
    \includegraphics[width=0.8\textwidth]{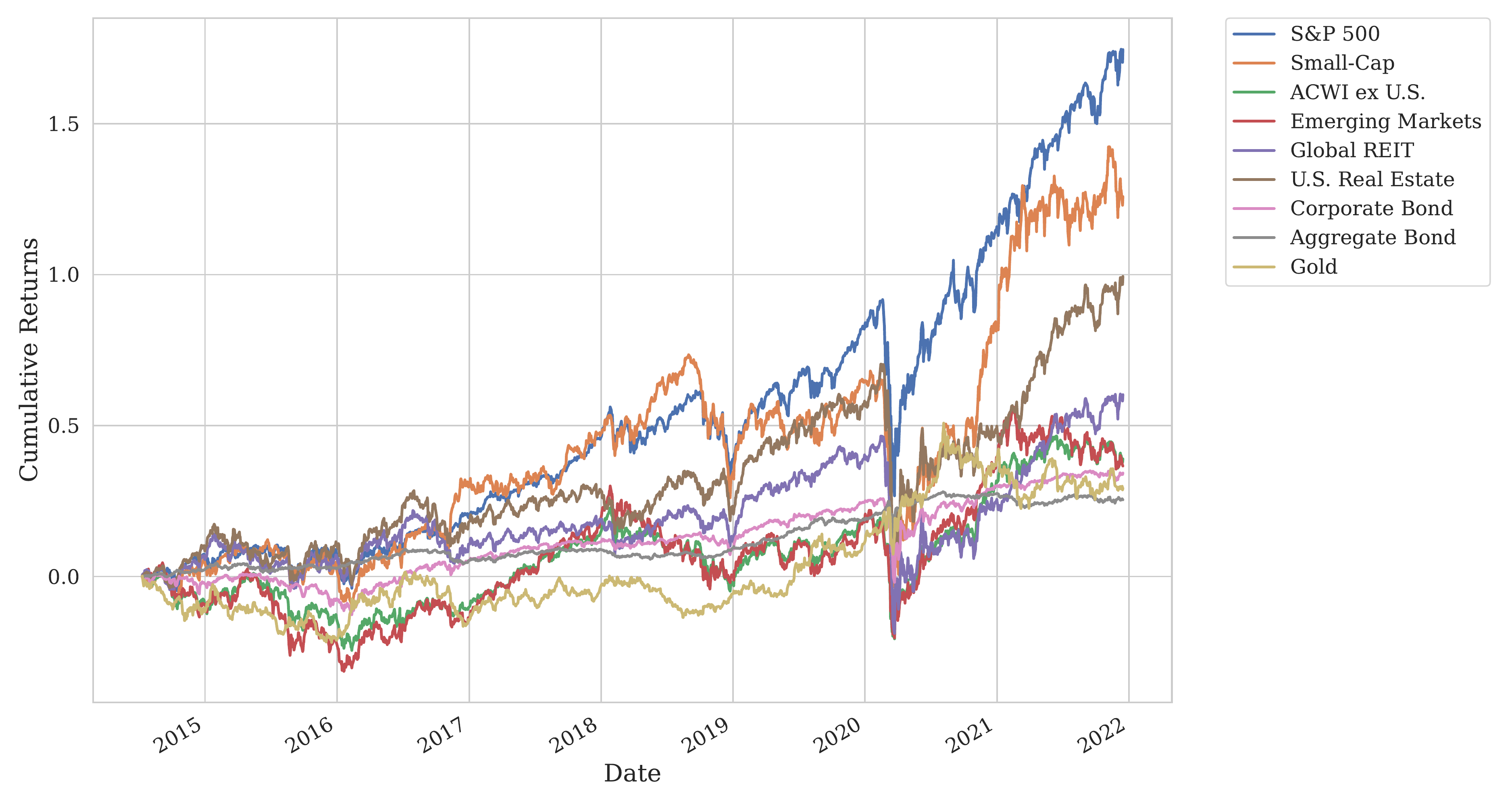}
    \caption{Cumulative Returns of Assets. Real Estate and U.S. equities have the highest returns in our data, while fixed-income assets were less volatile and with lower returns.}
    \label{fig:market_cum_returns}
\end{figure*}

We can now review the historical data according to the backtesting setup. We show the cumulative returns of the assets in Figure~\ref{fig:market_cum_returns}. As we expect, equities have higher returns than fixed incomes in general. 

\begin{figure*}[!t]
    \centering
    \includegraphics[width=0.7\textwidth]{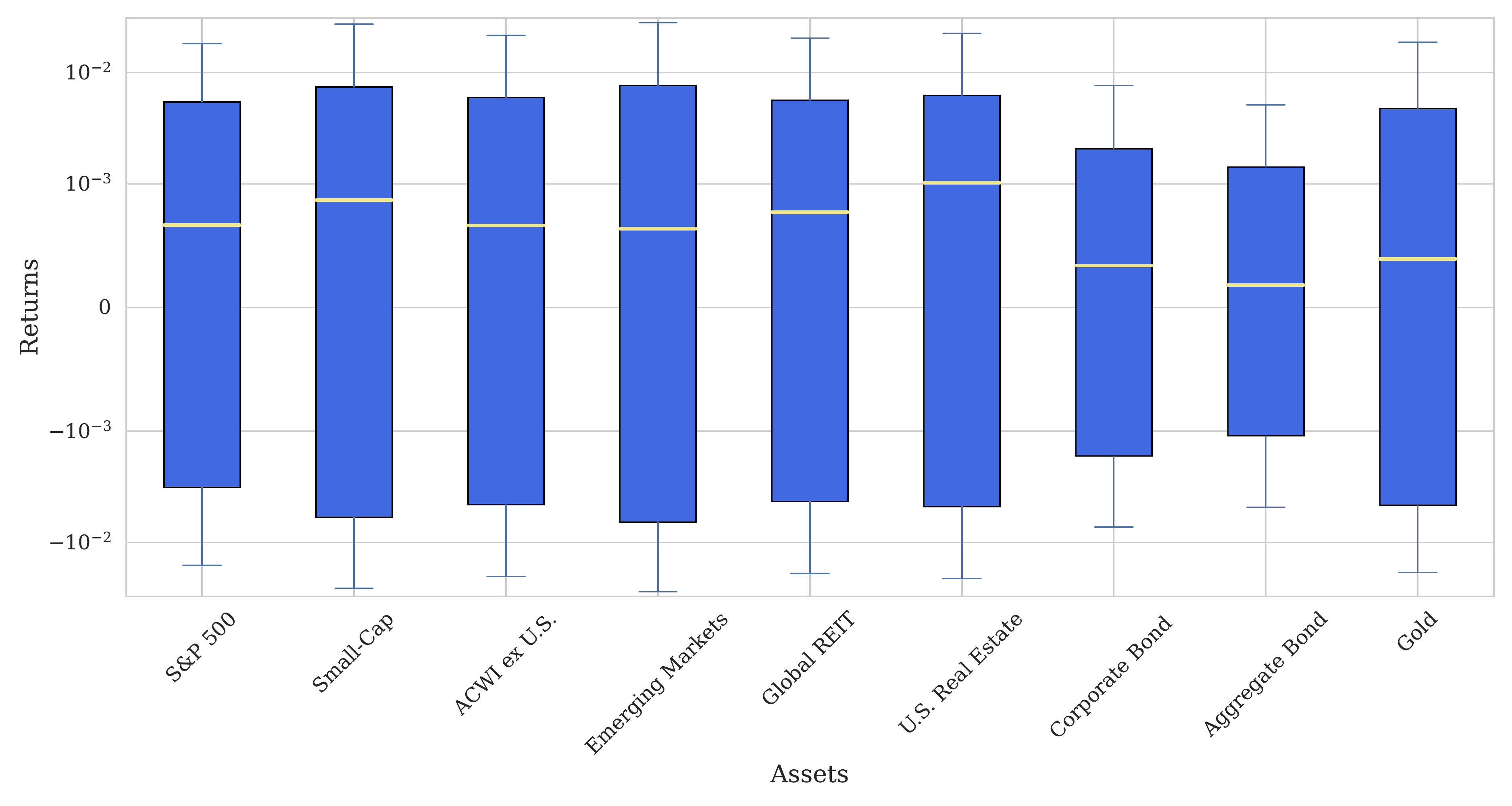}
    \caption{Distribution of Daily Asset Returns. Equities and gold are more volatile than fixed income. Equities' returns are higher than both gold and fixed income assets.}
    \label{fig:market_dist_returns}
\end{figure*}

In Figure~\ref{fig:market_dist_returns}, we illustrate the distribution of daily returns as box plots in symmetric log scale. 
Confirming the previous figure, this figure shows that the risks of equities and gold are more than fixed incomes.

\begin{figure*}[!t]
    \centering
    \includegraphics[width=0.8\textwidth]{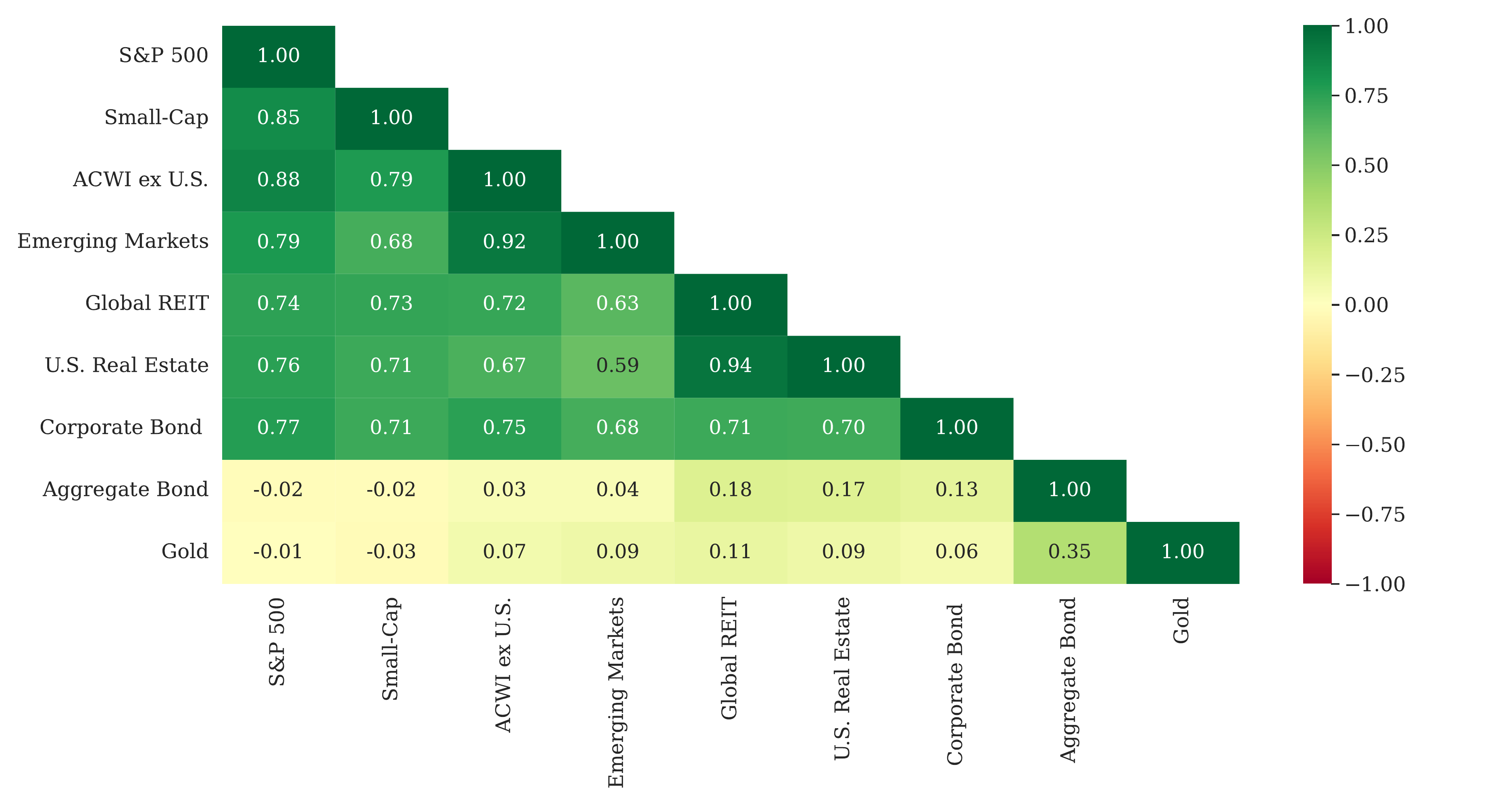}
    \caption{Correlation of Daily Asset Returns. Aggregate bond an gold have low correlation with the other assets, which makes them proper candidates to bring diversity to portfolio.}
    \label{fig:market_corr}
\end{figure*}

The correlations of daily returns reveal information about how the asset returns change compared to each other during the analysis period. The asset correlation in Figure~\ref{fig:market_corr} illustrates that equities, real estate, and corporate bonds are highly correlated, while aggregate bonds and gold are much less correlated with the rest of assets.

\subsection{Portfolios' Performance}
We use \toolname to create portfolio with different goals using the list of assets in Table~\ref{tab:assetlist} and configurations in Table~\ref{tab:backtestingparams}. We first introduce out-of-sample evaluation results.

\begin{figure*}[!t]
    \centering
    \includegraphics[width=0.8\textwidth]{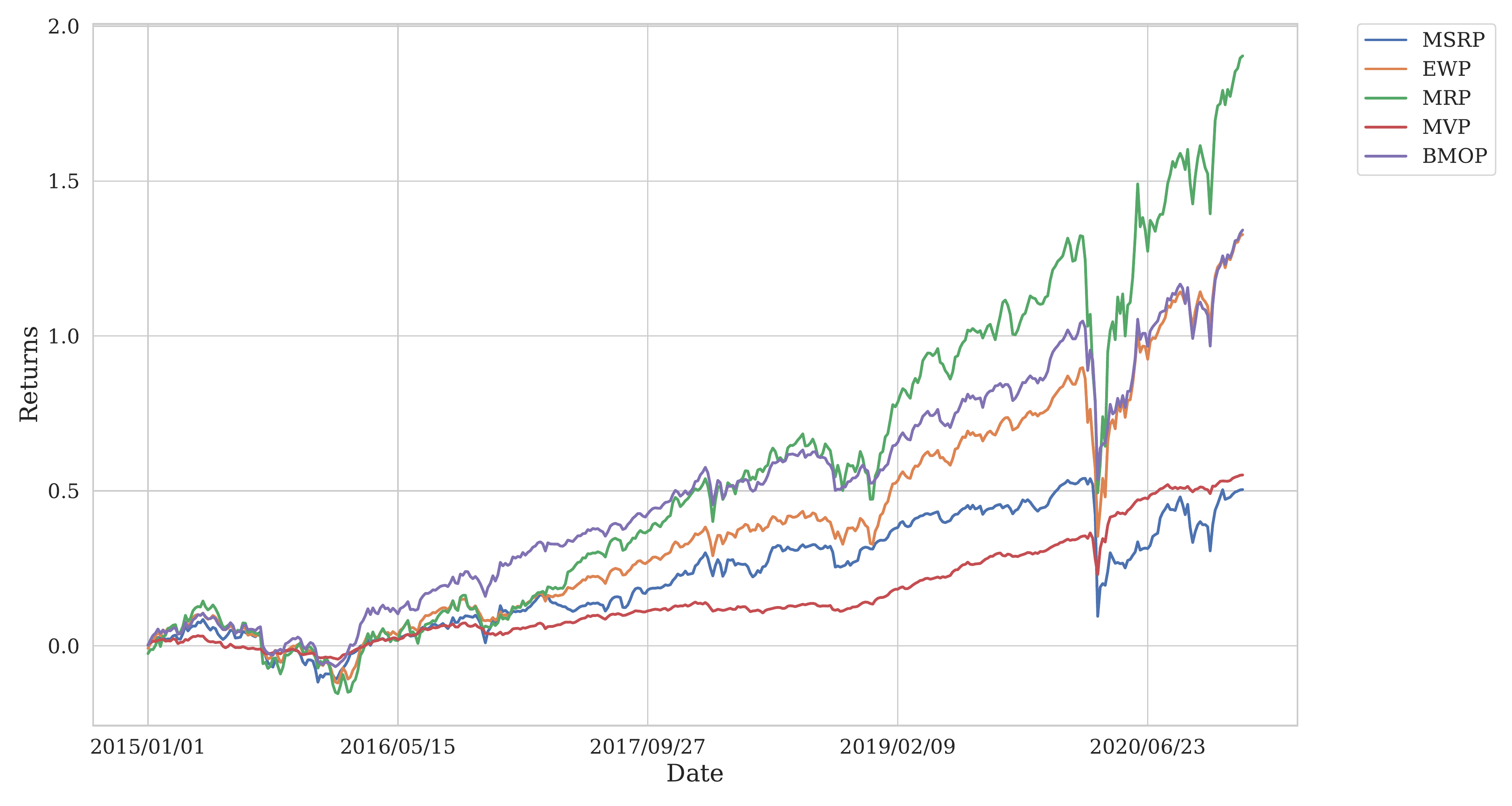}
    \caption{Cumulative Portfolio Returns. MRP and BMOP have the highest returns.}
    \label{fig:portfolio_cum_returns}
\end{figure*}

\begin{figure*}[!t]
    \centering
    \includegraphics[width=0.7\textwidth]{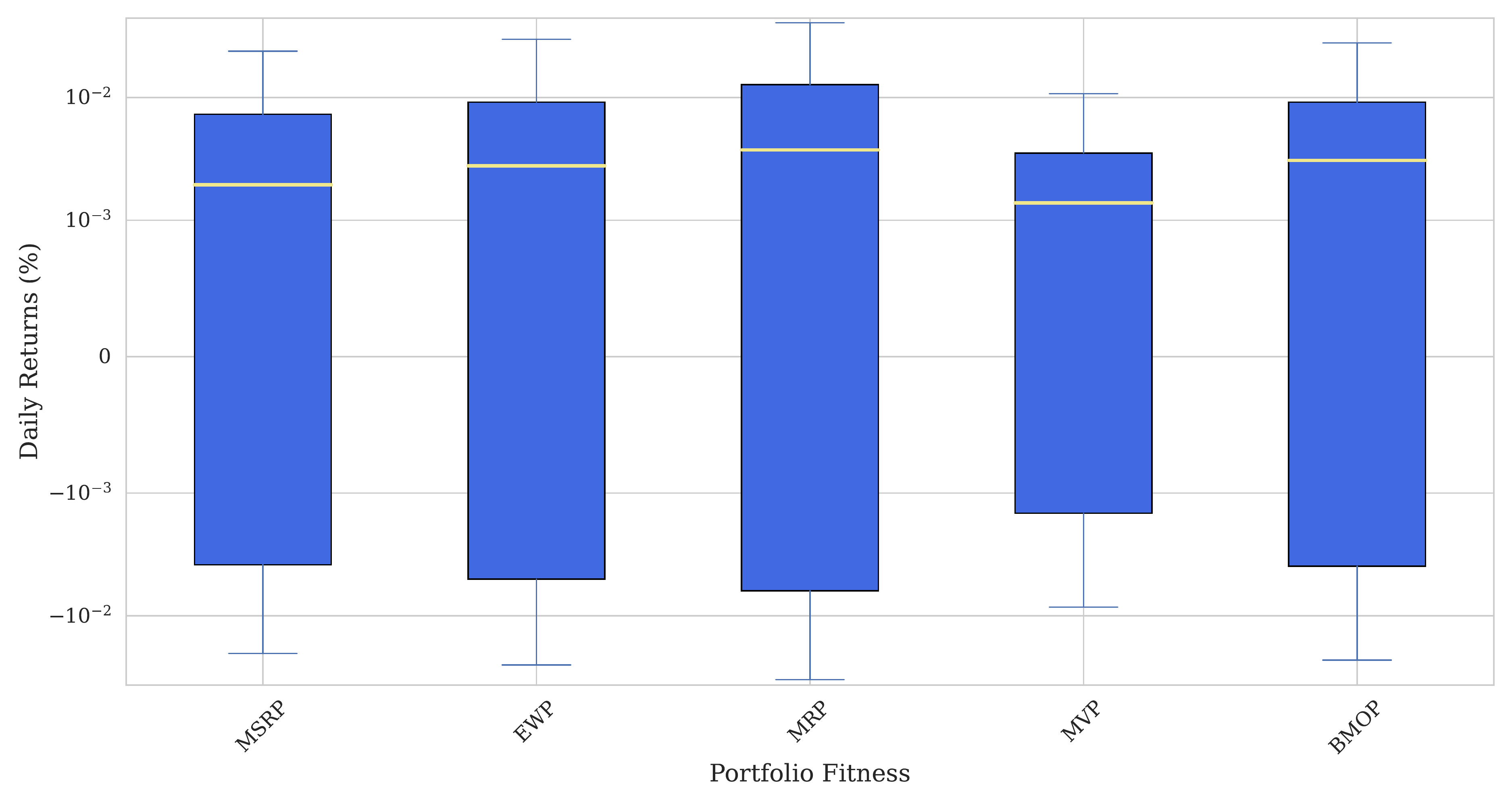}
    \caption{Distribution of Daily Portfolio Returns.}
    \label{fig:portfolio_dist_returns}
\end{figure*}

In Figure~\ref{fig:portfolio_cum_returns}, we illustrate the cumulative returns of a baseline portfolio (EWP), three mean-variance portfolios created on a classical computer (MRP, MVP, and MSRP), and a binary-weight portfolio (BMOP) created on a quantum computer.
The distributions of daily returns are also demonstrated in Figure~\ref{fig:portfolio_dist_returns}. As we expect, MRP has the highest returns while MVP is less volatile than the other portfolios, and MSRP seems a balance between MVP and MRP. While BMOP is formulated differently and cannot have fractional asset weights, it works very well which shows potential for using quantum computers in portfolio optimization. Furthermore, EWP seems an interesting one here since both its return and volatility are acceptable while it brings simplicity to the investment. 

\begin{figure*}[!t]
    \centering
    \includegraphics[width=0.7\textwidth]{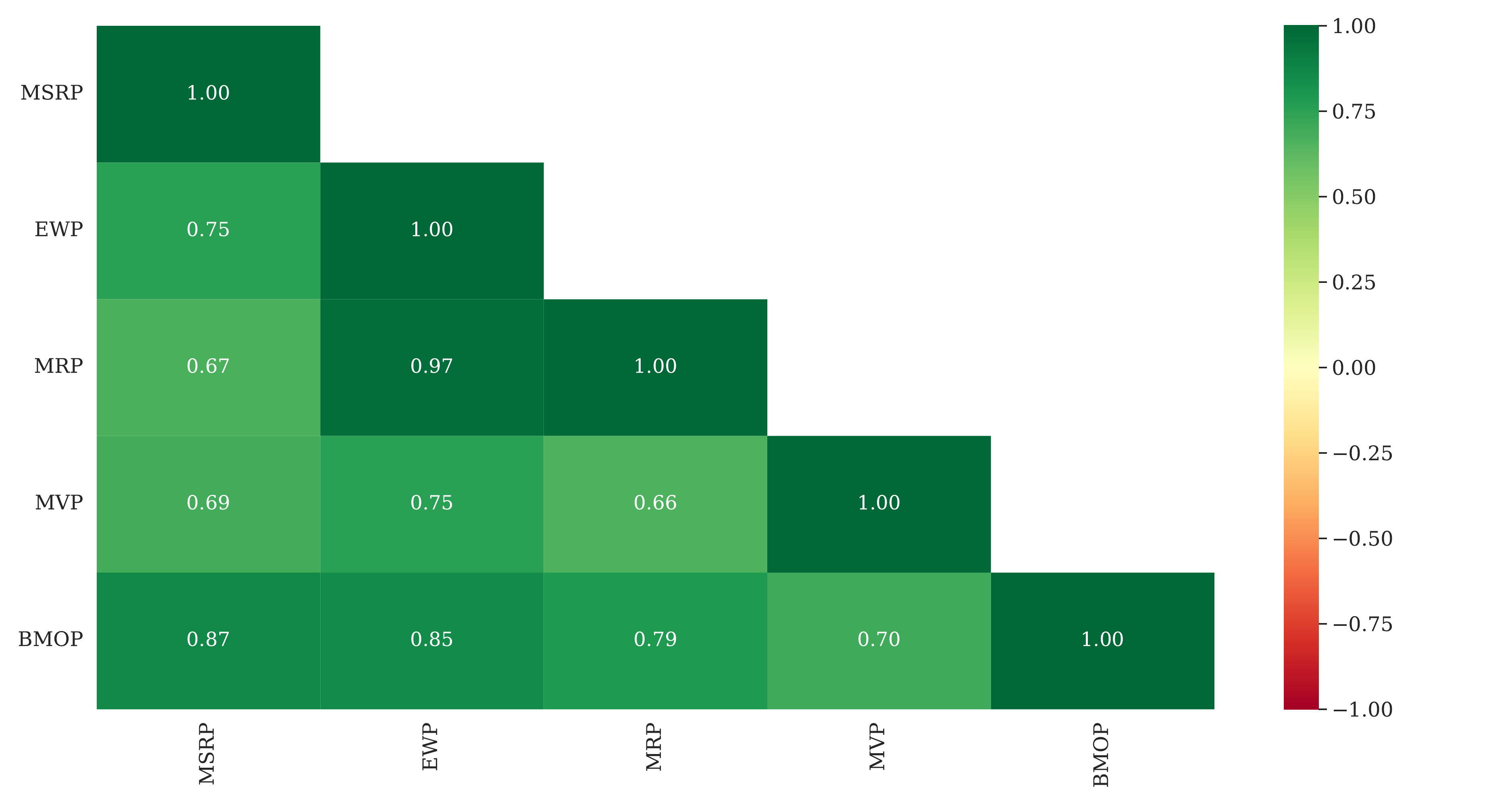}
    \caption{Correlation of Daily Portfolio Returns. MSRP and MVP are less correlated to the other portoflios.}
    \label{fig:portoflio_corr}
\end{figure*}

Figure~\ref{fig:portoflio_corr} shows the correlation between different portfolios. We can see a relatively high correlation between different approaches. 

\begin{figure*}[!t]
    \centering
    \includegraphics[width=0.8\textwidth]{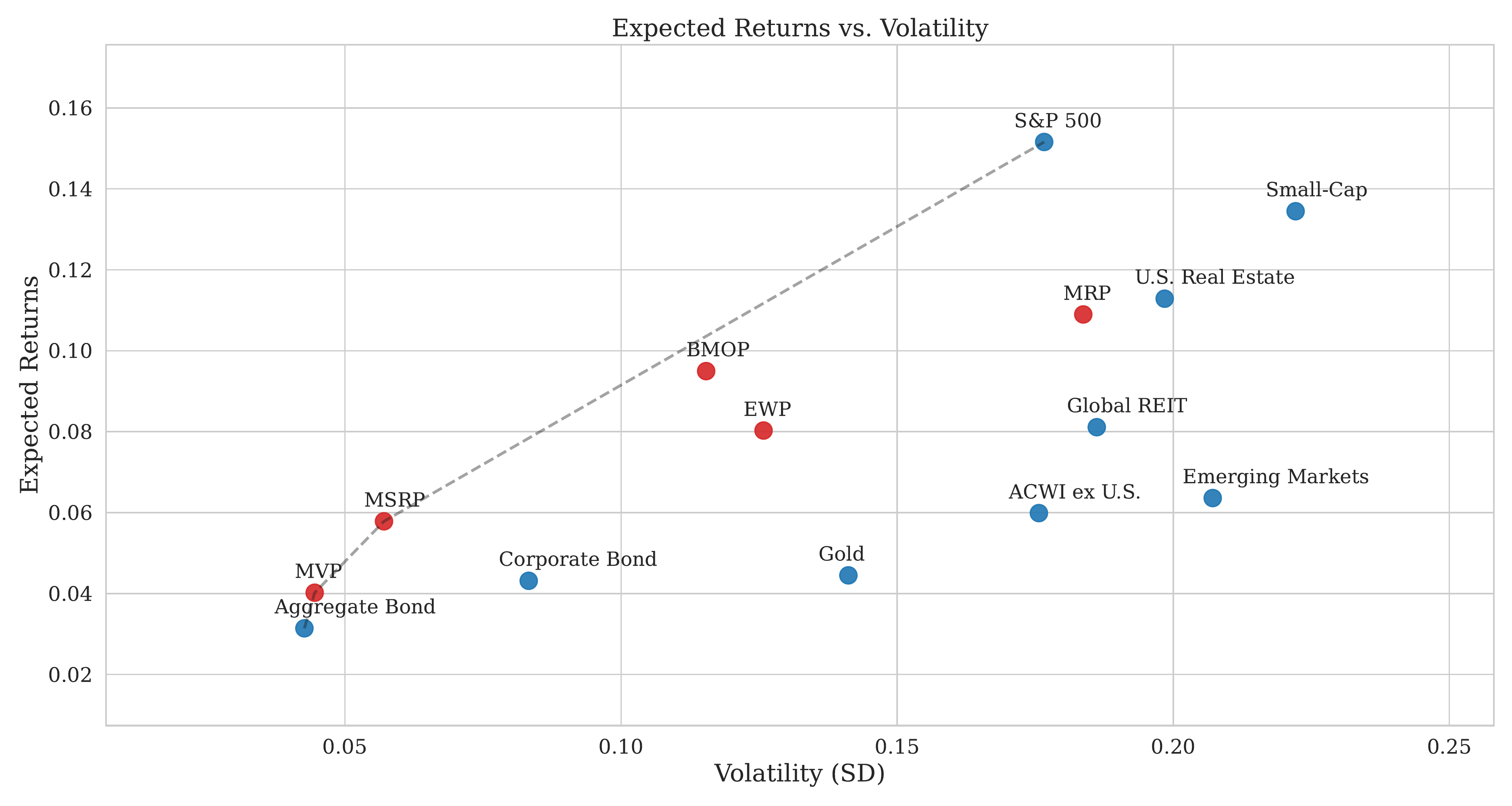}
    \caption{Expected returns vs. Volatility for assets (in blue) and portfolios (in red). Each set of asset weights results in a portfolio than can be mapped into this two-dimensional space. The ones on the efficient frontiers (the dashed black line) are the portfolios with minimum volatility for a given target expected return.}
    \label{fig:portfolio_multi}
\end{figure*}

To be able to compare the asset and portfolio returns and risks more clearly, we illustrate annualized volatility (standard deviation of returns) and expected (average) returns of assets (in blue) and portfolios (in red) in Figure~\ref{fig:portfolio_multi} as an in-sample evaluation. The dashed black line shows the efficient frontier. In general, investing in portfolios reduces volatility while sacrificing returns. Choosing a balance between these two depends on investors' preferences.

\section{Limitations of the Mean-Variance Portfolio} \label{sec:limitatons}

Despite the theoretical power of the mean-variance portfolio, the original form of this approach makes several unrealistic assumptions that limit its usage in practice. 
These assumptions can be relaxed by altering the cost function and constraints, but we focus only on the original form of the portfolio here. 

\paragraph{Asset Liquidity} The mean-variance portfolio assumes that all assets in the investment universe are completely liquid, and we can buy and sell them at any time. 
While the soft version of this assumption holds for large-cap assets in developed markets, it is not always the case for the rest of the assets. 

\paragraph{Costs} Turn-overs, transactions, and management are costly. 
However, the mean-variance portfolio does not consider these costs. 
Such a significant elimination could result in high returns and low risks in backtesting while having a poor performance in practice. 

\paragraph{Statistical Limitations} The mean-variance portfolio assumes that returns are independent and identically distributed (i.i.d), and they can be expressed solely by their returns and covariance matrix. 
Although this assumption could hold for a short period and some assets, its relevance needs to verify for specific historical data. 
This approach also assumes that the mean and variance of returns are the only factors an investor may aim to optimize and investors have a clear view of their real expectations about the returns they seek and the volatility they can tolerate.
Lastly, this approach assumes that the expected returns are a good prediction for future returns, which is not always the case.

\paragraph{Dividends} The mean-variance portfolio only considers price changes to calculate returns and eliminate dividends in calculations. 
This makes calculations unrealistic for high-dividend assets.  

\paragraph{Asset Concentration} The mean-variance portfolio approach tends to generate portfolios with a high concentration in only a few assets. This might be the local (or even global) optimal asset allocation according to the optimization problem definition, but it is not favourable from an investment viewpoint.

\section{Conclusion}

We introduce \toolname in this paper. This open-source Python library is capable of generating different mean-variance portfolio optimization scenarios according to the input parameters. 
Accepting a vast range of configuration parameters allows the end-user to overcome some of the restrictions and limitations of the original version of this approach to create more realistic portfolios. 
\toolname performs the whole pipeline necessary to create, evaluate, and verify portfolios, including data collection, optimization, backtesting, and visualization.
Our sample investment scenario demonstrates that \toolname can generate optimal portfolios for various targets.
We release \toolname code together with its documentation and highly encourage the reader to contribute to them.

\bibliographystyle{IEEEtran}
\bibliography{main}

\end{document}